\documentclass[epj-spec]{svjour}
\usepackage{graphics}

\newcommand{\be}{\begin{equation}}
\newcommand{\ee}{\end{equation}}
\newcommand{\bq}{\begin{eqnarray}}
\newcommand{\eq}{\end{eqnarray}}

\newcommand{\ket}[1]{\left | \, #1 \right\rangle}

\begin{document}
\title{Zero modes of various graphene configurations from the index theorem}
\author{Jiannis K. Pachos \inst{1}
\fnmsep\thanks{\email{j.k.pachos@leeds.ac.uk}} \and Agapitos Hatzinikitas \inst{2}
\and Michael Stone \inst{3} }
\institute{School of Physics and Astronomy, University of Leeds, Leeds LS2
9JT, UK
\and
Department of Statistics and Actuarial-Financial Mathematics, School of
Sciences, University of Aegean, 83200 Samos, Greece
\and
Department of Physics, University of Illinois, 1110 W. Green St. Urbana, IL
61801, USA}
\abstract{
In this article we consider a graphene sheet that is folded in various
compact geometries with arbitrary topology described by a certain genus,
$g$. While the Hamiltonian of these systems is defined on a lattice one can
take the continuous limit. The obtained Dirac-like Hamiltonian describes
well the low energy modes of the initial system. Starting from first
principles we derive an index theorem that corresponds to this Hamiltonian.
This theorem relates the zero energy modes of the graphene sheet with the
topology of the compact lattice. For $g=0$ and $g=1$ these results coincide
with the analytical and numerical studies performed for fullerene molecules
and carbon nanotubes while for higher values of $g$ they give predictions
for more complicated molecules. }
%
\maketitle
\section{Introduction}
\label{intro}

The spectrum of graphene and its various geometrical configurations has been
the focus of extensive study
\cite{DiVincenzo,Tworzydlo,Gonzalez1,Gonzalez,Lammert,Kolesnikov}.
It provides a physical system where a unique interplay is witnessed between
geometry and electronic properties such as conductivity. Nevertheless, a
unified picture has not been derived so far due to the richness in behavior
of the various geometrical configurations as well as the difficulty in
approaching them analytically. One of the interests is to study the number
of electronic eigenstates with zero energy that determine the conductivity
of the system and its ground state degeneracy. Previous methods for
obtaining the zero modes of the system are based on lengthy analytical or
numerical procedures. As a possible alternative the much celebrated index
theorem~\cite{Atiyah} offers an analytic tool that relates the zero modes of
elliptic operators with the geometry of the manifold on which these
operators are defined. This theorem has a dramatic impact on theoretical and
applied sciences~\cite{Eguchi}. It provides information about the spectrum
of widely used elliptic operators based on simple geometric considerations
that could be otherwise hard or even impossible to determine.

In this article we would like to describe the effect geometrical
deformations have on the spectrum of graphene. For that we shall establish a
version of the index theorem~\cite{Atiyah,Eguchi,Stone} that relates the
number of zero modes of graphene wrapped on arbitrary compact surfaces to
the topology of the surface. In our pursue we shall ignore changes in the
couplings caused by the geometrical deformations and we shall focus only on
the effect the geometry has on the spectrum of graphene. As we shall see our
results are in good agreement with the known cases of icosahedral fullerene
molecules~\cite{Kroto} and graphite nanotubes~\cite{Reich} where the
spectrum has been determined analytically or numerically. Similar approaches
for the ground state degeneracy of fractional quantum Hall systems in the
planar case or on high-genus Riemannian surfaces have been taken
in~\cite{Semenoff:1984dq,Schakel:1990mv,Jackiw1,Wen,Alimohammadi}.

\section{The graphene sheet}
\label{Graphene}

First we shall present an overview of the properties of a flat sheet of
graphene. When considering its low energy limit a linearization of the
energy is possible due to the presence of individual Fermi points in the
spectrum. This results in a Dirac equation~\cite{Gonzalez1}, which describes
well the low energy behavior of the system.

Graphene consists of a two dimensional honeycomb lattice where Carbon atoms
occupy its vertices. When we adopt the tight-binding approximation the model
reduces the system of coupled fermions on a honeycomb
lattice~\cite{Gonzalez} (see Fig.~\ref{lattice}). The relevant Hamiltonian
is given by
\begin{equation}
H = -J\sum_{<i,j>}a_i^\dagger a_j,
\label{Ham}
\ee
where $J>0$ denotes the tunneling coupling of the electrons along the
lattice sites, $<\!\! i,j\!\!>$ denotes nearest neighbors and $a^\dagger_i$,
$a_i$ are the fermionic creation and annihilation operators at site $i$ with
the non-zero anticommutation relation $\{a_i,a_j^\dagger\}=\delta_{ij}$. The
original lattice can be split into two triangular sublattices, A and B, that
correspond to the black and blank circles in Fig.~\ref{lattice}. This
facilitates the evaluation of the dispersion relation of graphene, which is
given by
\be
E(p) = \pm J\sqrt{1+ 4 \cos^2{\sqrt{3} p_y \over 2} + 4 \cos {3 p_x
\over 2}\cos {\sqrt{3} p_y \over 2} },
\label{dispersion}
\ee
where the distance between lattice sites is normalized to one. By solving
the equation $E(p)=0$ one deduces that, at half-filling, graphene possesses
two independent Fermi points, denoted by ${\bf K}_+$ and ${\bf K}_-$,
instead of Fermi lines. This rather unique property makes it possible to
linearize its energy by expanding it near the conical singularities of the
Fermi points. It is not hard to show that by restricting near the Fermi
points the resulting Hamiltonian takes the form of the Dirac operator
\be
H_{\pm}= \pm{3 J\over 2} \gamma^\alpha p_\alpha,
\label{Dirac}
\ee
where repeated indices are summed over the spatial cooridiantes $x,y$. The
Dirac matrices, $\gamma^\alpha$, are given by the Pauli matrices,
$\gamma^\alpha=\sigma^\alpha$, and $\pm$ corresponds to the two independent
and oppositely positioned Fermi points. Hence, the low energy limit of
graphene is described by a free fermion theory. Eigenstates of this Dirac
operator are two dimensional vectors, called spinors, given by $(\ket{{\bf
K}_\pm A},\ket{{\bf K}_\pm B})^T$, where A and B denote the two sublattices
and ${\bf K_\pm}$ denote two independent Fermi points chosen such that ${\bf
K}_-=-{\bf K_+}$.
\begin{center}
\begin{figure}[ht]
\resizebox{!}{3 cm}
{\includegraphics{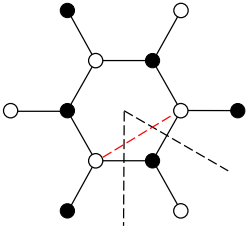} }
\caption{\label{lattice} The honeycomb lattice comprises of two
triangular lattices, A, denoted by black circles and, B, denoted by blank
circles. A single pentagonal deformation can be introduced by cutting a $\pi
/3$ sector and gluing the opposite sites together.}
\end{figure}
\end{center}

\section{Curvature deformations and effective gauge fields}
\label{Curvature_deformations}

Here we are interested in surfaces with arbitrary topology so we need to
introduce curvature in the initially flat honeycomb lattice. This is
achieved by selectively inserting lattice deformations. In doing so, we
shall demand that each lattice site has exactly three neighbors and that the
lattice is inextensional that is it is free to bend, but impossible to
stretch. The minimal alteration of the honeycomb lattice that can introduce
curvature without destroying the cardinality of the sites is the insertion
of a pentagon or a heptagon; this corresponds to locally inserting positive
or negative curvature, respectively. Other geometries are also possible,
leading to similar results as we shall see in the following.

To introduce a single pentagon in a honeycomb lattice, one can cut a $\pi/3$
sector and glue the opposite sides together, as illustrated in
Fig.~\ref{lattice}. This causes no other defects in the lattice structure.
We shall demand that the spinors are smooth along the cut remedied by
introducing compensating fields which negate the
discontinuity~\cite{Lammert,Kolesnikov}. Indeed, the cut introduced in
Fig.~\ref{lattice} causes an exchange between A and B sublattices. This
discontinuity can be remedied by introducing in the Hamiltonian the
non-abelian gauge field $\mathbf{A}$ circulation
\be
\oint A_\mu dx^\mu = {\pi \over 2} \tau_y
\nonumber
\ee
where $\tau_y$ is the Pauli operator that mixes the ${\bf K}_+$ and ${\bf
K}_-$ spinor components. This flux can be attributed to a fictitious
magnetic monopole inside the surface with a charge contribution of $1/8$ for
each pentagon~\cite{Coleman}. In addition, moving a frame around the
pentagonal deformation gives a non-trivial coordinate transformation. The
effect of this transformation on the spinors can be described by a spin
connection ${\bf \Omega}$. This is chosen such that its flux around the
pentagon is given by
\be
\oint \Omega_\mu dx^\mu = -{\pi \over 6}\sigma_z
\nonumber
\ee
and measures the angular deficit of $\pi/3$ around the cone.

The modified Dirac equation, which incorporates the curvature and the
effective gauge field, couples the ${\bf K}_\pm$ spinor components together
due to the non-abelian character of ${\bf A}$. Since this is the only mixing
term they can be decoupled by a single rotation that gives
\be
{3J \over 2} \gamma^\mu(p _\mu -i\Omega_\mu -i A_\mu^k)\psi^k =E\psi^k,
\label{Dirac2}
\ee
where $k=1,2$ denotes the components in the rotated basis with the
circulation of the abelian now field given by $\oint A^k_\mu dx^\mu
=\pm\pi/2$ (no summation is considered in $k$). The curved space Dirac
matrixes $\gamma^\mu$ are given by $\gamma^\mu=\sigma^\alpha e^\mu_\alpha$,
where $e^\mu_\alpha$ is the zweibein of the curved surface with metric
$g_{\mu
\nu}$ that defines the local flat reference frame,
$\eta_{\alpha\beta}=e^\mu_\alpha e^\nu _\beta g_{\mu\nu}$. They satisfy the
anti-commutation relations $\{\gamma^\mu,\gamma^\nu\} =2g^{\mu \nu}$, where
$g^{\mu \nu}$ is the inverse of $g_{\mu \nu}$. The curvature of the surface
is given by the tensor
$$
{{\mathcal R}^\mu}_{\nu\rho\sigma} = \partial_\sigma \Gamma^\mu_{\nu \rho}
-\partial_\rho
\Gamma^\mu_{\nu \sigma} +\Gamma^\lambda_{\nu \rho}
\Gamma^\mu_{\lambda\sigma} -\Gamma^\lambda_{\nu \sigma}
\Gamma^\mu_{\lambda\rho}
$$
where the Christoffel symbols are defined by
$$
\Gamma^\sigma_{\mu \nu} = {1 \over 2} g^{\sigma \rho} (\partial_\mu g_{\nu
\rho} +\partial_\nu g_{\mu \rho} -\partial_\rho g_{\mu \nu})
$$
The Ricci tensor is given by ${\mathcal R}_{\mu \nu} \equiv {{\mathcal
R}^\sigma}_{\mu \nu\sigma}$ and the scalar curvature is given by ${\mathcal
R}\equiv g^{\mu \nu} {\mathcal R}_{\mu\nu}$. The field strength that
corresponds to the abelian gauge potential, $A_\mu^k$, is given by
${\mathcal F}^k_{\mu\nu} =\partial_\mu A_\nu^k -
\partial_\nu A_\mu^k$. Equation (\ref{Dirac2}) faithfully describes the low
energy behavior of graphene, such as its zero modes, when it is deformed to
an arbitrary surface.

\section{Index theorem and graphene}
\label{Application}

\subsection{The index theorem}
\label{Index_Theorem}

Since the obtained Dirac operator is an elliptic operator, it is possible to
employ the index theorem~\cite{Atiyah,Eguchi,Stone} to gain information
about its low energy spectrum. Indeed, the index theorem gives an insight in
the structure of the spectrum of certain operators without the need to
diagonalize them. This information can be derived from general properties of
the operators and the geometry of the space, $M$, they are defined on. A two
dimensional Dirac operator defined on a surface coupled to a gauge field can
be given by the general form
$$
\slash\!\!\!\! D =\left(
\begin{array}{cc} 0 & P^\dagger \\ P &0
\end{array}\right)
$$
where $P$ is an operator that maps from a space $V_+$ to the space $V_-$,
while $P^\dagger$ maps from $V_-$ to $V_+$. As we are interested in the zero
modes, we can define the dimension of the null subspace of $P$ and
$P^\dagger$ by $\nu_+$ and and $\nu_-$ respectively. To facilitate the
bookkeeping we introduce the chirality operator $\gamma_5$ by
$$
\gamma_5 =\left(
\begin{array}{cc} 1 & 0 \\ 0 & -1
\end{array}\right) = \sigma_z
$$
so that it anticommutes with the Dirac operator and it has the states in
$V_\pm$ as eigenstates with corresponding eigenvalue $\pm 1$. As we are
interested in the zero modes we can consider the operator $\slash\!\!\!\!
D^2$, which is diagonal
$$
{\slash\!\!\!\! D} ^2 =\left(
\begin{array}{cc} P^\dagger P & 0 \\ 0 & P P^\dagger
\end{array}\right)
$$
and has the same number of zero modes as $\slash\!\!\!\! D$. One can easily
show that the operators $PP^\dagger$ and $P^\dagger P$ have the same number
of non-zero eigenstates. Indeed, if there is a state $u$ such that
$PP^\dagger u=\lambda u$ then the state $P^\dagger u$ is an eigenstate of
the operator $P^\dagger P$ with the same eigenvalue, $P^\dagger P (P^\dagger
u) =\lambda (P^\dagger u)$. In order to demonstrate the index theorem we
shall employ the heat kernel expansion method \cite{Vassilevich}. Consider a
two dimensional compact surface, $M$. Then one can consider the expansion
\be
Tr (\hat f e^{-t\hat D}) = {1 \over 4\pi t} \sum_{k\geq 0} t^{k/2} a_k(\hat
f,\hat D)
\label{kernel}
\ee
where $Tr$ denotes the trace of matrices and the integration of spatial
coordinates. $a_n(\hat f,\hat D)$ are the expansion coefficients that one
needs to determine as a function of the operators $\hat f$ and $\hat D$. For
$\hat f=\gamma_5$ and $\hat D=\slash\!\!\!\! D^2$ one deduces that
$$
Tr(\gamma_5 e^{-t\slash \!\!\!\!D^2} ) = Tr (e^{-t P^\dagger P}) - Tr (e^{-t
P P^\dagger}) = \sum_{(\lambda_1)} (e^{-t \lambda_1}) - \sum_{(\lambda_2)}
(e^{-t \lambda_2})
$$
where $\lambda_1$ and $\lambda_2$ are the eigenvalues of the operators
$P^\dagger P$ or $PP^\dagger$ respectively. But we have shown that for every
eigenstate of the operator $P^\dagger P$ there is a corresponding eigenstate
of $PP^\dagger$ with exactly the same eigenvalue. Thus only the zero
eigenvalues remain giving
$$
Tr(\gamma_5 e^{-t\slash \!\!\!\!D^2} ) =\nu_+ - \nu_-
$$
Combining this result with relation (\ref{kernel}) it is possible to deduce,
that, for $f=\gamma_5$ and $D=\slash\!\!\!\! D^2$, all of the coefficients
$a_k$ should be zero except for $a_2$ where $a_2=\nu_+ -
\nu_-\equiv index(\slash\!\!\!\! D)$. The value of $a_2$ defined from
(\ref{kernel}) can be found from the first order term in the $t$ expansion
of the exponential. Considering that $\slash \!\!\!\!D^2 = -g^{\mu \nu}
\nabla_\mu \nabla_\nu + {i \over 4}[\gamma^\mu, \gamma^\nu] {\mathcal F}_{\mu \nu} -{1
\over 4} {\mathcal R}$, where $\nabla_\mu$ is the reparametrization and gauge covariant
derivative, one can easily deduce that
$$
a_2 = Tr \left[\gamma_5 ({i \over 4} [\gamma^\mu,\gamma^\nu] {\mathcal
F}_{\mu\nu} -{1 \over 4} {\mathcal R})\right]= 2 \int \!\!\!\!\int {\mathcal
F}
$$
where ${\mathcal F}$ is the field strength, ${\mathcal R}$ is the scalar
curvature and the integration runs over the whole compact surface $M$. Thus,
we have an analytic way to evaluate the index of $\slash\!\!\!\! D$ by
\be
index(\slash\!\!\!\! D) ={1 \over 2\pi} \int\!\!\!\!\int {\mathcal F}
\label{index_th}
\ee
The absence of the curvature term in this formula is due to the traceless
nature of $\gamma_5$ and it is a characteristic of two dimensions. If
one can evaluate the integral of the field strength over the whole compact
surface then the least number of zero modes is determined. It is worth
noting that for compact surfaces this integral is an integer due to the
Dirac quantization condition of the monopole charges \cite{Coleman}.

\subsection{Application to graphene}
\label{Application_Graphene}

Our aim is to evaluate the contribution from the gauge field, ${\mathcal
F}$, in (\ref{index_th}) for the particular case of a folded sheet of
graphene in a compact surface. In a previous section we determined how for
each lattice deformation a gauge field circulation is introduced. If one
could determine the total number of deformations for a particular compact
geometry of the lattice then we would be able to determine the
$index(\slash\!\!\!\! D)$. At this point we shall assume that curvature is
introduced by only inserting pentagons and heptagons in the lattice.
Interestingly, one can evaluate the number of such deformations in a lattice
necessary to generate a compact surface by employing the Euler
characteristic. Indeed, for $V$, $E$ and $F$ being respectively the number
of vertices, edges and faces of a lattice defined on a compact surface with
genus $g$, the Euler characteristic, $\chi$, is given by
\be
\chi = V - E + F = 2(1-g).
\nonumber
\ee
Take the total number of pentagons, hexagons and heptagons in the lattice to
be, $n_5$, $n_6$ and $n_7$, respectively. Then the total number of edges is
given by $E = (5n_5 +6n_6 + 7n_7)/2$ as each polygon $n_i$ contributes $i$
edges, but each edge is shared by two polygons. Similarly the total number
of vertices and faces can be evaluated to be $V= (5n_5 +6n_6 + 7n_7)/3$ and
$F=n_5 +n_6 + n_7$, giving finally
\be
n_5-n_7=12(1-g).
\label{Euler}
\ee
This result signifies that non-trivial topologies necessarily introduce an
imbalance in the numbers of pentagons and heptagons. Moreover, inserting
equal numbers of pentagons and heptagons do not change the topology of the
surface as they cancel out. This is consistent with the effective gauge
field description where a pentagon and a heptagon have opposite flux
contributions. As particular examples we see that Eqn. (\ref{Euler})
reproduces the known case of a sphere with $g=0$ giving $\chi=2$ and a
number of defects $n_5=12$ and $n_7=0$. This is the lattice of the C$_{60}$
fullerene. For the torus we have $g=1$ for which $\chi=0$ and $n_5=n_7=0$
reproducing the lattice of the nanotubes. For a genus-2 surface we have
$\chi=-2$, $n_5=0$, $n_7=12$. In all these examples equal numbers of
pentagons and heptagons can be inserted without changing the topology of the
surface.

Now we are in position to evaluate the $index(\slash\!\!\!\!D)$. The
contribution of the gauge field term in~(\ref{index_th}) can be calculated
straightaway from the Euler characteristic. It is obtained by adding up the
contributions from the surplus of pentagons or heptagons. Thus, the total
flux of the effective gauge field can be evaluated by employing Stokes's
theorem, giving
\be
{1 \over 2\pi} \int \!\!\!\!\int\! {\mathcal F} ={1 \over
2\pi}\sum_{n_5-n_7}\! \oint A= {1\over 2\pi} (\pm{\pi\over 2}) (n_5-n_7)=
\pm 3(1-g),
\nonumber
\ee
where the sign $\pm$ corresponds to the $k=1,2$ gauge fields and the
summation runs over all the surplus of pentagons or heptagons. Hence, from
(\ref{index_th}), one obtains
\be
index(\slash\!\!\!\!D)= \nu_+-\nu_- = \left\{\begin{array}{cc}
\,\,\,\,\,3(1-g),& \textrm{for} \,\, k=1\\ -3(1-g),& \textrm{for}
\,\, k=2 \end{array}\right..
\label{the_index}
\ee
Consequently, the least number of zero modes is given by $6|1-g|$, which
coincides with their exact number if $\nu_-=0$ or $\nu_+=0$. This is
actually what happens in most of the cases when the index is non-zero.

\subsection{Zero modes for fullerenes and nanotubes}
\label{Examples}

The above result relates the number of zero modes of a graphene sheet with
the genus of the surface it has been folded. As expected it reproduces the
number of zero modes for the known molecules. The fullerene, for which genus
$g=0$, has six zero modes which correspond to the two triplets of C$_{60}$
and of similar larger molecules~\cite{Gonzalez,Samuel}. For the case of
nanotubes, we consider periodic boundary conditions, which give effectively
a torus with $g=1$. In this case, formula (\ref{the_index}) gives
$\nu_+-\nu_-=0$. This is in agreement with previous theoretical and
experimental results \cite{Saito,Reich} which have the nanotubes with either
no zero modes (e.g. zigzag nanotubes) or with zero modes that satisfy
$\nu_+=\nu_-$ (e.g. armchair nanotubes). This is a consequence of the
symmetry between the two opposite directions along the nanotube.

\section{Conclusions}
\label{Conclusions}

In this article we have employed the index theorem~\cite{Atiyah} to
enumerate the zero modes of a graphene sheet when compactified on arbitrary
genus surfaces. Our results are in good agreement with the presently studied
cases of fullerenes and nanotubes. The only approximation employed here was
the continuous limit for obtaining the Dirac operator. This approximation is
valid if we restrict to the low energy spectrum of the system described by
large wavelengths. In this limit, the lattice spacing or the conical
singularities of the pentagonal deformations do not affect the low energy
modes. Thus, larger fullerene molecules than the C$_{60}$ provide more
accurate results.

As an additional example we can consider a graphene sheet folded on an
octahedron. In this case six square plaquettes have to be inserted in the
honeycomb lattice. As square plaquettes do not have the frustration
properties of pentagons, no effective gauge field is introduced. Thus, the
index in this case is zero, agreeing with previous
considerations~\cite{Gonzalez}. Beyond the known examples the version of the
index theorem presented here gives a relation between the zero modes of more
complex molecules. Even if the latter are defined on compact surfaces they
are related to experimentally relevant ones by imposing appropriate periodic
conditions. Thus, the metallic properties of these molecules can be induced
by simple geometrical considerations.

{\acknowledgement} This research was supported by the Royal Society.

\end{document}